\documentclass{JHEP}
\usepackage{amsmath,amssymb,graphicx,xspace}
\newcommand{\CFTvacuum}{\ensuremath{|0\rangle}\xspace}
\newcommand{\CFTvacuumeta}{\ensuremath{|0\rangle_{\eta}}\xspace}
\newcommand{\BTZvacuum}{\ensuremath{|M,J\rangle}\xspace}
\newcommand{\BTZvacuumeta}{\ensuremath{|M,J\rangle_{\eta}}\xspace}

\title{Lodged in the throat:\\
 Internal infinities and AdS/CFT }
\author{D. Marolf \\ Physics Department, UCSB, Santa Barbara, CA 93106 \\ marolf@physics.ucsb.edu}
\author{A. Yarom \\ Department of Physics, Ben-Gurion University, Beer-Sheva 84105 \\  yarom@bgumail.bgu.ac.il}

\abstract{In the context of AdS${}_3$/CFT${}_2$, we address
spacetimes with a certain sort of internal infinity as typified by
the extreme BTZ black hole.  The internal infinity is a null circle
lying at the end of the black hole's infinite throat. We argue that
such spacetimes may be described by a product CFT of the form
CFT${}_L \otimes$CFT${}_R$, where CFT${}_R$ is associated with the
asymptotically AdS boundary while CFT${}_L$ is associated with the
null circle.  Our particular calculations analyze the CFT dual of
the extreme BTZ black hole in a linear toy model of
AdS${}_3$/CFT${}_2$. Since the BTZ black hole is a quotient of
AdS${}_3$, the dual CFT state is a corresponding quotient of the CFT
vacuum state.  This state turns out to live in the aforementioned
product CFT.   We discuss this result in the context of general
issues of AdS/CFT duality and entanglement entropy.}


\preprint{hep-th/0511225}

\begin{document}

\section{Introduction}
\label{S:introduction}

The anti-de Sitter/conformal field theory correspondence (AdS/CFT)
\cite{MaldacenaAdSCFT} is a powerful tool that has shed light on
many interesting aspects of physics (see e.g. \cite{AdSCFTreview}),
and especially that of black holes.  In particular, it has
elucidated calculations of black hole entropy in string theory (e.g.
\cite{StromVafa}), and has provided strong motivation for the idea
that black hole evaporation should be a unitary
process\footnote{See, however \cite{Ted} for a contrasting view.}.

However, fundamental questions concerning the degrees of freedom
associated with black holes remain unanswered. For example, we still
lack a bulk calculation of black hole entropy in terms of
microstates. Another issue of interest in the context of AdS/CFT is
just what CFT should be used to describe the most general black hole
geometries.  Classical gravity can describe black holes with a
variety of complicated interiors such as those containing inflating
universes or a second asymptotic region. One notes that such
examples seem to require additional degrees of freedom beyond the
CFT (which we shall call CFT${}_0$) used to describe AdS space
itself \cite{HM1, BKLT,BGC,Maldacena:2001,inflate,KOS}.
Unfortunately, we remain far from a general understanding of the
CFT's in which states dual to such geometries might live.

Here we describe another such scenario.  We argue below that, in a
certain context, the dual description of an {\em extreme} black hole
may require additional degrees of freedom beyond those of CFT${}_0$.
Although it lacks a second asymptotic region, the extreme black hole
has an internal infinity lying at the end of its infinite throat
\cite{throat}. As we discuss briefly below, such an internal
infinity can also be considered as part of the boundary of the bulk
spacetime, and can provide a home for these additional degrees of
freedom. The internal infinity is marked $\hat D_L$ on the conformal
diagram shown in figure \ref{F:conf}.

\FIGURE[p]{
\scalebox{.5}{\includegraphics{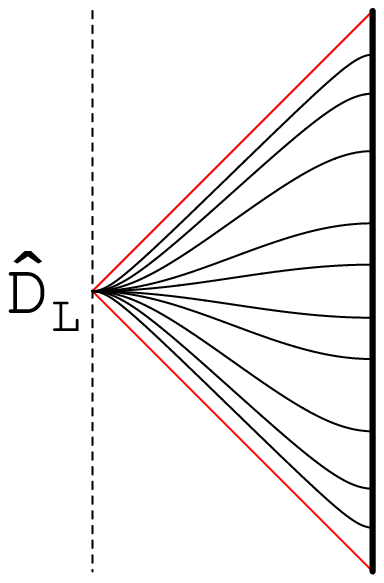}}%
\label{F:conf}%
\caption{A conformal diagram of the extreme BTZ black hole, with
lines of constant BTZ coordinate $t$ (see equation
\ref{E:BTZ_line_element}) drawn for later reference.  Each point on
the diagram represents an $S^1$ in the spacetime which is the orbit
of a spacelike Killing field.  The diagonal lines are the horizon,
the heavy line at right is the conformal boundary at $r = \infty$,
the dashed line is the singularity, and the point $\hat D_L$ marks
the internal infinity which forms the `end' of the infinite throat.}
}

Now, the reader may be concerned by the fact that extreme black
holes are often described within CFT${}_0$
\cite{MaldacenaAdSCFT,AdSCFTreview}, without the addition of any new
degrees of freedom. To avoid confusion, let us point out that one
may consider two distinct classes of spacetimes containing extreme
black holes: those with an infinite throat (which we address in this
paper) and those without. The usual eternal extreme black hole
(figure \ref{F:conf}) is clearly an example of the first class, as
is any spacetime generated from it by sending in small perturbations
from its asymptotic boundary.

The other class of spacetimes arises when extreme black holes form
dynamically.  Of course, this cannot happen by any classical
process.  Consider, however, a nearly extreme black hole with one
asymptotic region (perhaps formed from the collapse of matter). As a
result of a thermal fluctuation, such a black hole may decay to
extremality, emitting some Hawking radiation in the process. In the
semiclassical description, the decay occurs because of a negative
flux of energy across the future horizon. Thus one may expect that,
before some advanced time $v$, the spacetime is that of a
non-extreme black hole.  Thus, it has no infinite throat.

A similar dichotomy arises when one compares non-extreme black holes
with differing numbers of asymptotic regions (i.e., one vs. two). In
that case, one expects the number of such regions to be reflected by
differing dual CFTs \cite{HM1,BKLT,BGC,Maldacena:2001}. It is
natural to expect that our two classes of spacetimes, with and
without extreme black hole internal infinities, should correspond to
two distinct CFTs as well.  The class without an infinite throat
(but with one asymptotic region) should be described by CFT${}_0$,
while, due to the additional boundary conditions needed at the
internal infinity, the class with an infinite throat should be
described by a larger CFT.

It is of course possible that the infinite throat is simply a red
herring (e.g., as suggested in
\cite{Senblackhole,Mathur,MathurReview} and references therein).
However, pushing this model forward may provide insight into the
broader issues of black holes and dual degrees of freedom. We are
also interested in the relation to entanglement entropy\footnote{
Bulk discussions of entanglement entropy have been of interest for
some time \cite{Israel,Bombelli,Srednicki,
Jacobson,KabStr,Jacobson:1994iw,Frolov:1998vs,
Larsen:1995ax,Kabat:1995eq,Holzheyetal,
Casini:2003ix,AreaScaling,Cramer:2005mx}, though several issues
remain unclear. These include the species problem (see e.g.
\cite{WaldRev}), the correct value of the cut-off used in
entanglement entropy calculations (see \cite{Sorkin,width}), and
other related issues (see, e.g., \cite{triologue}).} in the context
of AdS/CFT \cite{Maldacena:2001,HawMalStr,entinST}. We therefore
investigate features associated with the throat of the extreme BTZ
black hole below.

Our approach will be to use a simple linear toy model of
AdS${}_3$/CFT${}_2$, which was considered implicitly in
\cite{MalStr} and then more explicitly in \cite{LouMar}.  The
model replaces the CFT${}_2$ of \cite{MaldacenaAdSCFT} by a single
real-valued free scalar field on the cylinder. Empty AdS${}_3$ is
 of course taken to be dual to the vacuum of this CFT. As we remind the
reader in section \ref{S:review}, the BTZ black hole
\cite{BTZ1,BTZ2} can be constructed as a quotient AdS${}_3/\Gamma$
of AdS${}_3$, where $\Gamma$ is an appropriately acting discrete
group. The boundary of the BTZ black hole is an analogous quotient
of the boundary $\partial$AdS${}_3$ of AdS${}_3$.  Since the model
CFT is linear, there is a natural map which takes the CFT state on
the (boundary) spacetime $S^1\times\mathbb{R}$ and constructs an
associated CFT state on the quotient boundary
$S^1\times\mathbb{R}/\Gamma$.

In the non-extreme case, the appropriate quotient construction leads
to a black hole with two asymptotic regions, and thus with two
asymptotic boundaries, each of which is identical to the boundary of
pure AdS${}_3$. Here we will reexamine this construction in detail,
focussing on the extreme limit.  One asymptotic boundary, which we
take to be the right boundary, remains intact and is again identical
to the boundary of pure AdS${}_3$.  Thus, it is natural for a copy
of the original CFT to be associated with this boundary.  We refer
to this copy as CFT${}_R$. Although the second asymptotic region
disappears in this limit, we will nevertheless find that the state
dual to an extreme black hole lives in a product conformal field
theory, CFT${}_L\otimes$CFT${}_R$, where CFT${}_L$ is associated
with the `end' of the infinite throat of the extreme black hole.
This internal infinity is a remnant of the second asymptotic
boundary of the non-extreme black hole  which, as we shall see
below, has degenerated to a null circle.  As a result, CFT${}_L$ has
only right-moving degrees of freedom.

The plan of this paper is as follows: Section \ref{S:review}
provides a brief review of the BTZ black hole and sets notation
for the rest of this paper. In section \ref{S:vacuum} we adopt the
method of \cite{LouMar} to obtain the dual description of the BTZ
black hole for all masses and angular momenta. In particular,
section \ref{S:extremal} elaborates on the extreme BTZ black hole.
Finally, we discuss the implications of our results for AdS/CFT in
section \ref{S:discussion}.

\section{Review: the BTZ black hole}
\label{S:review}

Recall \cite{BTZ1,BTZ2} that the BTZ black hole is a solution to 2+1
dimensional gravity with negative cosmological constant. Outside the
horizon, the line element of this solution is given by
\begin{equation}
\label{E:BTZ_line_element}
    ds^2=-\frac{(r^2-r_+^2)(r^2-r_-^2)}{r^2 \ell^2}dt^2
    + \frac{r^2\ell^2}{(r^2-r_+^2)(r^2-r_-^2)}dr^2
    + r^2\left(d\phi+\frac{r_+r_-}{r^2\ell }dt\right)^2,
\end{equation}
from which one notes the presence of Killing horizons at
$r_+,r_-$. In the same notation, the mass of the black hole is
\begin{align}
\label{E:BTZmass}
    M&=\frac{r_+^2+r_-^2}{8 \ell^2 G_{(3)}},\\
\intertext{and the angular momentum is}%
\label{E:BTZangmom}
    J&=\frac{r_+ r_-}{4 \ell G_{(3)}},
\end{align}
where $G_{(3)}$ is the three-dimensional gravitational constant.
Here $r_+ \ge r_- \ge 0$ and $\ell$ is the AdS length scale in the
sense of equation (\ref{E:AdSinR22}) below.

Since gravity has no local degrees of freedom in 2+1 dimensions,
the BTZ solution is locally just AdS${}_3$. In fact, the BTZ
solution may be thought of as the quotient of a certain region in
AdS${}_3$ by an appropriate discrete group of isometries. Section
\ref{S:BTZ} reviews the bulk aspects of this quotient
construction. In section \ref{S:boundary} we review the quotient
of the conformal boundary $\partial$AdS${}_3$ of AdS${}_3$ which
will give $\partial$BTZ, the boundary of our BTZ black hole. In
both subsections our main focus is a proper description of the
extreme limit $r_+ \rightarrow r_-$.

\subsection{BTZ as a quotient}
\label{S:BTZ}

We are interested in the description of the BTZ black hole
(\ref{E:BTZ_line_element}) as a quotient of AdS$\phantom{}_3$
\cite{BTZ2}. We remind the reader that AdS$\phantom{}_3$ is the
universal covering space of the surface $\widehat{\rm
AdS}\phantom{}_3$ defined by the relation
\begin{equation}
\label{E:AdSinR22}
    -\ell^2 = -(T^1)^2-(T^2)^2+(X^1)^2+(X^2)^2
\end{equation}
in $\mathbb{R}^{2,2}$  with line element:
\begin{equation}
\label{E:R22metric}
    ds^2=-(dT^1)^2-(dT^2)^2+(dX^1)^2+(dX^2)^2.
\end{equation}

We will use the coordinates $(\rho,t,\theta)$ adopted in
\cite{AdS3orbifoldanalysis},
\begin{align}
\label{E:embeddingT1}
    T^1&=\ell\frac{1+\rho^2}{1-\rho^2} \sin t,
    &T^2=\ell\frac{1+\rho^2}{1-\rho^2} \cos t\\
    X^1&=\ell\frac{2\rho}{1-\rho^2} \cos \theta,
\label{E:embeddingX2}
    &X^2=\ell\frac{2\rho}{1-\rho^2} \sin \theta
\end{align}
with $0<\rho<1$ and  $-\pi<\theta \le \pi$.  For $\widehat{\rm
AdS}\phantom{}_3$ we have $-\pi \le t < \pi$, but we are
interested in the universal cover, AdS${}_3$, which has $-\infty <
t < \infty$. In such coordinates, the line element becomes
\begin{equation}
\label{E:AdS3global}
    ds^2=\frac{4 \ell^2}{1-\rho^2}\left(
        -\frac{1}{4}(1+\rho^2)^2 dt^2 + d\rho^2 +\rho^2 d\theta^2
        \right).
\end{equation}
After the conformal rescaling
\begin{equation}
d\tilde s^2 = \left(\frac{1-\rho^2}{4 \ell^2}\right) ds^2,
\end{equation}
the boundary ($\rho \to 1$) line element is just that of the
standard cylinder
\begin{equation}
\label{E:AdS3boundary}
    d\tilde s_{\partial AdS_3}^2=-dt^2+d\theta^2.
\end{equation}

The six generators of the $SO(2,2)$ isometries of
AdS$\phantom{}_3$ are given by
\begin{equation}
\label{E:SO22generators}
    J_{ab}=x_b \frac{\partial}{\partial x_a}-x_a
    \frac{\partial}{\partial x_b}
\end{equation}
with $x^{a} \in (T^1,T^2,X^1,X^2)$.  Consider the Killing vector
\begin{equation}
\label{E:Killingspace}
    \xi=\frac{r_+}{\ell}J_{T^2,X^1}-\frac{r_-}{\ell}J_{T^1,X^2}+J_{X^1,X^2}-J_{T^2,X^2}
\end{equation}
with $r_+ \geq r_- \geq 0$. The BTZ black hole is obtained by
identifying points $P$ in AdS$\phantom{}_3$ along the orbits of
$\xi$ at intervals of Killing parameter $2\pi n$, with $n \in
\mathbb{Z}$:
\begin{equation}
\label{E:identification}
    P \sim e^{2\pi \xi n} P.
\end{equation}

By applying this quotient procedure to the region $\xi^2>0$ we
obtain a global description of the BTZ black hole. The coordinate
transformation relating the coordinates of
(\ref{E:BTZ_line_element}) to (\ref{E:AdS3global}) on the quotient
space can be found in \cite{BTZ2,MalStr}. The geometry of the
resulting quotient depends only on the conjugacy class of $\xi$
within SO(2,2). If $r_+ \neq r_-$, one may choose a representative
$\xi'$ of the conjugacy class of (\ref{E:Killingspace}) such that
\begin{equation}
\label{E:Killingspace_2}
    \xi'=\frac{r_+}{\ell}J_{T^2,X^1}-\frac{r_-}{\ell}J_{T^1,X^2}.
\end{equation}
An explicit coordinate transformation which takes
(\ref{E:Killingspace}) into $(\ref{E:Killingspace_2})$ for $r_+
\neq r_-$ can be found in \cite{BTZ2}. The simpler form
(\ref{E:Killingspace_2}) has the property that $\xi \rightarrow
-\xi$ under an inversion of the space directions: $(X_1,X_2)
\rightarrow (-X_1,-X_2)$.  As a result, a quotient construction
based on (\ref{E:Killingspace_2}) is manifestly symmetric under
this inversion. However, the representation
(\ref{E:Killingspace_2})  is not possible for the extreme black
holes ($r_+ = r_-$) on which we wish to focus. As a result,  we
use (\ref{E:Killingspace}) instead of the simpler
(\ref{E:Killingspace_2}).  In doing so, we note that our
parametrization (\ref{E:Killingspace}) explicitly breaks the
$(X_1,X_2) \rightarrow (-X_1,-X_2)$ symmetry.

The identifications (\ref{E:identification}) act on the region
$\xi^2 > 0$ in AdS$\phantom{}_3$.  Other regions are not considered
as they would lead to closed causal curves or to singularities in
the quotient space: points in the bulk of AdS with $\xi^2=0$ project
onto what is termed the singularity of the black hole in
\cite{BTZ1,BTZ2}.

Although AdS${}_3$ is maximally symmetric, the BTZ black hole has
only two isometries. To identify them, we note that these descend
from the two Killing fields of AdS${}_3$ which commute with $\xi$.
One is $\xi$ itself (\ref{E:Killingspace}), which by construction
projects to a spacelike Killing field on the quotient. The second,
$\eta$, may be taken to be proportional to the lift of the time
translation symmetry of the BTZ black hole.  Let us denote the
projection of these Killing fields to the BTZ black hole by $\hat
\xi$ and $\hat \eta$. Comparing with (\ref{E:BTZ_line_element}), we
find that $\hat \xi = \frac{\partial}{\partial \phi}$ and $\hat \eta
= \ell \frac{\partial}{\partial t}$ on the BTZ spacetime, while on
AdS${}_3$ we find:
\begin{equation}
\label{E:def_of_eta}
    \eta=-J_{T^1,T^2}+J_{T^1,X^1}-\frac{r_-}{\ell} J_{T^2,X^1}+\frac{r_+}{\ell}
J_{T^1,X^2}.
\end{equation}

\subsection{The Quotient of the Boundary}
\label{S:boundary}

As noted in section \ref{S:introduction}, we will be especially
interested in the action of the quotient (\ref{E:identification})
on the boundary  $\partial{\rm AdS}_3$ of  ${\rm AdS}_3$. We now
study this action in detail.

It is convenient to introduce null coordinates $u = t+\theta$, and
$v= t - \theta$.  In terms of these coordinates, the  Killing
fields take the form
\begin{align}
\label{E:bKVFs1}
    \xi=&2\sqrt{1+\Sigma^2}\cos\left(\frac{u}{2}\right)\cos\left(\frac{u}{2}-\arctan\Sigma\right)
        \partial_u \\
        -&2\sqrt{1+\Delta^2}\cos\left(\frac{v}{2}\right)\cos\left(\frac{v}{2}+\arctan\Delta\right)
        \partial_v,\\
    \eta=&2\sqrt{1+\Sigma^2}\cos\left(\frac{u}{2}\right)\cos\left(\frac{u}{2}-\arctan\Sigma\right)
        \partial_u \\
        +&2\sqrt{1+\Delta^2}\cos\left(\frac{v}{2}\right)\cos\left(\frac{v}{2}+\arctan\Delta\right)
        \partial_v,
\label{E:bKVFs2}
\end{align}
on $\partial{\rm AdS}_3$, where we have defined $\Sigma =
\frac{r_+ + r_-}{\ell}$ and $\Delta =\frac{ r_+ - r_-}{\ell}$.

We wish to identify the region in  $\partial$AdS${}_3$ where
$\xi^2\ge 0$, as only this region will project to the boundary of
the BTZ black hole.  From (\ref{E:bKVFs1}) we have
\begin{equation}
    \xi^2 = 4\sqrt{(1+\Delta^2)(1+\Sigma^2)}
    \cos\left(\frac{u}{2}\right)\cos\left(\frac{v}{2}\right)
    \cos\left(\frac{u}{2}-\arctan \Sigma\right)\cos\left(\frac{v}{2}+\arctan
    \Delta\right).
\end{equation}
 It is convenient to write the region with $\xi^2>0$ as
$D_R \cup D_L$, where
\begin{align}
\label{E:diamonds} D_R &= \{-\pi+2 \arctan \Sigma < u < \pi,\,-\pi
< v < \pi - 2\arctan
    \Delta\}\\
D_L &=\{\pi < u <\pi+2\arctan \Sigma,\,\pi-2\arctan \Delta < v <
\pi \},
\end{align}
together with the images of $D_R \cup D_L$ under the translations
$u \rightarrow u + 2\pi n$ and $v \rightarrow v + 2\pi n$.   For
$r_+ \neq r_-$, the quotients $\hat{D}_R$ of $D_R$ and $\hat{D}_L$
of $D_L$ form the respective conformal boundaries of the left and
right asymptotic regions of the BTZ black hole.

\FIGURE[p]{
\scalebox{1}{\includegraphics{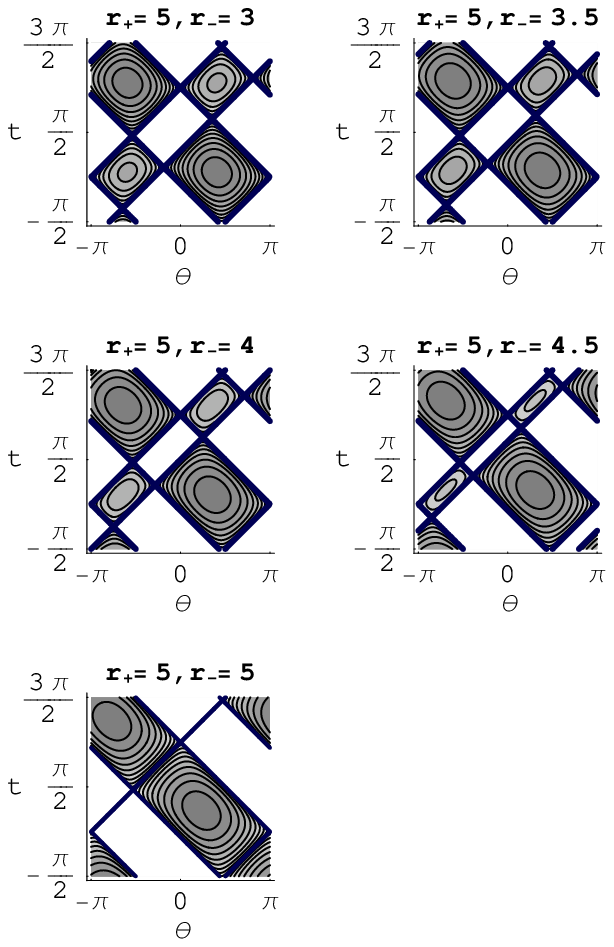}}%
\label{F:xib_deq1_toextreme}%
\caption{A contour plot of $\xi^2$ on the boundary of AdS${}_3$,
which under the identification $P\sim e^{2\pi n \xi}P$ becomes the
boundary of the BTZ black hole. Regions with $\xi^2 > 0$ are shaded,
with darker shading for higher values of $\xi^2$. Negative values of
$\xi^2$ are shaded white. The thick line marks where $\xi^2=0$. As
the black hole becomes extremal, one boundary collapses to a null
line.}
}

It is interesting to note that $D_L$ and $D_R$ do not appear
symmetrically in (\ref{E:diamonds}).  Indeed, $D_R$ is a diamond
of coordinate dimensions $(\pi - 2 \arctan \Sigma)\times(\pi -
2\arctan \Delta)$, while $D_L$ is a diamond of coordinate
dimensions $(2 \arctan \Sigma) \times (2 \arctan \Delta)$. For
$r_+ \neq r_-$, this is a result of our choice of
(\ref{E:Killingspace}) over (\ref{E:Killingspace_2}) and the
explicit breaking of the symmetry $(X_1,X_2) \rightarrow
(-X_1,-X_2)$. On the other hand, the lack of symmetry is no
surprise for extreme black holes (for which $r_+=r_-$ so that
$\Delta=0$), as such black holes have only one asymptotic region.

A plot of the regions $D_R$ and $D_L$ (in $(t,\theta)$ coordinates)
is given in figure \ref{F:xib_deq1_toextreme}. One clearly sees
that, as the black hole approaches extremality, the left diamond
$D_L$ collapses to a null line. Despite the fact that $\xi^2=0$ on
this null line, it is convenient to still refer to it as $D_L$.
Under the quotient (\ref{E:identification}) $D_L$ maps to a null
circle $\hat{D}_L$. For $M=0$, $D_L$ degenerates to a point, which
is in fact a fixed point of (\ref{E:identification}).

For extreme black holes with $M \neq 0$, we will find in section
\ref{S:vacuum} below that interesting degrees of freedom live on
the null circle $\hat{D}_L$.  As a result, we would like to think
of it as part of the boundary of the black hole.  In the conformal
compactification of AdS${}_3$, the null line $D_L$ forms part of
the boundary of the region with $\xi^2 > 0$. Curves approaching
$D_L$ project to curves which travel down the infinite throat of
the extreme BTZ black hole. Thus, we may think of this null circle
as lying at the end of the throat.

Now, it is clear that $D_L$ will not form part of the smooth
conformal boundary of the BTZ black hole.  However, (see figure
\ref{F:conf}) there are both spacelike and causal curves (e.g., the
generators of the extreme BTZ horizon) which reach $D_L$ from within
the $\xi^2 > 0$ region. As a result, one should be able to use
causal boundary techniques (e.g, \cite{MRCB}, which builds on
\cite{CB1,CB2,CB3,CB4,CB5,CB6,CB7,CB8,CB9})
to give a rigorous sense in which this null circle forms part of the
BTZ boundary and to establish in detail its relation to the infinite
throat.

\section{The dual of the BTZ black hole}
\label{S:vacuum}%

As stated in the introduction, our model of AdS/CFT is obtained by
replacing the CFT with a theory of a single minimally-coupled
massless free scalar field $\phi$ on $S^1 \times \mathbb{R}$. This
model theory is of course conformal, and has central charge $c=1$.
The advantage of this model is its linearity, so that the geometric
quotient construction of section \ref{S:review} has a natural
analogue in the CFT itself. For simplicity, we also replace the ten
dimensional bulk spacetime AdS${}_3 \times S^3 \times T^4$ with
AdS${}_3$. Here we follow \cite{LouMar} and, implicitly,
\cite{MalStr} which noted that this simple model is able to
reproduce a number of features of the full correspondence such as
the energy, angular momentum, and entropy, as well as the more
general thermal nature of the BTZ black hole.

In section \ref{S:review}, the BTZ black hole was described as the
quotient of AdS${}_3$ by a certain discrete group $\Gamma$, whose
action on AdS${}_3$ depends on the mass $M$ and the angular momentum
$J$ of the black hole.   Similarly, we found that the boundary
$\partial$BTZ of the BTZ black hole could be described as the
quotient of $D_L \cup D_R \subset \partial$AdS${}_3$ by the action
of $\Gamma$. Now, since the dual CFT is associated with the boundary
manifold, $\Gamma$ has a natural action on the CFT as well.  If
operators on $\partial$BTZ are identified with $\Gamma$-invariant
operators on $\partial$AdS${}_3$, the CFT state \CFTvacuum (dual to
empty AdS${}_3$) induces a state \BTZvacuum dual to the BTZ black
hole. The state \BTZvacuum is the part of \CFTvacuum which contains
information about those field modes which are periodic under the
identifications (\ref{E:identification}); information about the
other field modes is discarded\footnote{More precisely, we use the
fact that \CFTvacuum is a Gaussian state and we define \BTZvacuum to
be the Gaussian state on the boundary of the BTZ spacetime whose
covariance (equivalently, the two-point function of $\phi$ in this
state) is just the restriction of the covariance of \CFTvacuum to
those field modes on $D_L \cup D_R$ which are the lift of field
modes on the quotient.}. We will interpret \BTZvacuum as the CFT
state dual to the corresponding BTZ black hole.

After addressing the general case in section \ref{S:general}, we
highlight certain features of the extreme case in section
\ref{S:extremal}.

\subsection{The general case}
\label{S:general}

As in \cite{LouMar}, we shall begin by describing \CFTvacuum in
terms of the lift of modes which are positive frequency on
$\partial$BTZ. To do so, we seek solutions of the massless free wave
equation on $\partial$AdS${}_3$ which, when restricted to $D_L$ and
$D_R$, are positive frequency with respect to the BTZ time
translation symmetry $\eta$. To proceed we introduce  null
coordinates $\alpha$ and $\beta$ on the BTZ boundary and its
covering space $D_R \cup D_L$, where we require $\alpha$ and $\beta$
to satisfy $ \label{E:ab} \frac{\partial}{\partial \alpha} = \eta +
\xi$ , and $\frac{\partial}{\partial \beta} = \eta - \xi$.
This determines $\alpha$ and $\beta$ up to constants $\alpha_0$ and
$\beta_0$ in each diamond:
\begin{align}
\label{E:alpha}
    \alpha&=\frac{\ln\left(
        \left(\sqrt{1+\Sigma^2}\right)
        \frac{\cos\left(\frac{u}{2}-\arctan(\Sigma)\right)}{\cos\left(\frac{u}{2}\right)}\right)}{\Sigma}+\alpha_0,
            &\Sigma \neq 0.\\
\label{E:alpha2}
    \alpha&=\tan\left(\frac{u}{2}\right)+\alpha_0, &\Sigma = 0.\\
\label{E:beta}
    \beta&=-\frac{\ln\left(
        \left(\sqrt{1+\Delta^2}\right)
        \frac{\cos\left(\frac{v}{2}+\arctan(\Delta)\right)}{\cos\left(\frac{v}{2}\right)}\right)}{\Delta}+\beta_0,
            & \Delta \neq 0.\\
\label{E:beta2}
    \beta&=\tan\left(\frac{v}{2}\right)+\beta_0, &\Delta = 0.
\end{align}

We would like our coordinates to be real-valued. In $D_R$, the
sign of $\cos(u)$ is equal to the sign of
$\cos\left(\frac{u}{2}-\arctan \Sigma \right)$, so the argument of
the logarithm in (\ref{E:alpha}) is positive and we may take
$\alpha_0=0$. In contrast, in $D_L$, the argument of the logarithm
is negative and we may take $\alpha_0=-i \frac{\pi}{\Sigma}$,
where we have chosen the branch cut of the logarithm in
(\ref{E:alpha}) to be in the upper half $u$-plane. Similarly, for
$\beta_0$, we may take $\beta_0=0$ in $D_R$ and $\beta_0=-i
\frac{\pi}{\Sigma}$ in $D_L$.

One may use $u_{\omega,+}^R = \frac{1}{\sqrt{4 \pi \omega}} e^{-i
\omega \alpha}$ as a basis for right-moving solutions of the wave
equation in $D_R$ which are positive frequency with respect to
$\eta$, and similarly take $u_{\omega,+}^L=\frac{1}{\sqrt{4 \pi
\omega}}e^{i \omega \alpha}$ as a basis for right-moving solutions
of the wave equation in $D_L$ which are positive frequency with
respect to $\eta$.  Note that $\eta$ is future timelike in $D_R$ and
past timelike in $D_L$, while $\xi$ points to the right in $D_R$ and
to the left in $D_L$. In this notation it is easy to see that
$\overline{u_{\omega,+}^L}$ differs from the analytic continuation
of $u_{\omega,+}^R$ to $D_L$  by a factor of $e^{-\frac{\pi
\omega}{\Sigma}}$. A similar statement is true for the left-moving
modes $u_{\omega,-}^{L,R}$

Following \cite{Unruh}, we may use this observation to express
\CFTvacuum in terms of the state \CFTvacuumeta which is the
zero-particle state as defined by the modes $u^{R,L}_\pm$.  To do
so, note that for $\omega > 0$ the modes
\begin{align}
\label{E:W1p}
    W_{\omega,+}^{(1)}&=
        \frac{ e^{\frac{\pi \omega}{2 \Sigma}} u_{\omega,+}^R + e^{-\frac{\pi
        \omega}{2 \Sigma}} \overline{u_{\omega,+}^L}}{\sqrt{2\sinh\left(\frac{\pi\omega}{\Sigma}\right)}}
       ,
    &W_{\omega,+}^{(2)}=
        \frac{  e^{\frac{\pi \omega}{2 \Sigma}} u_{\omega,+}^L + e^{-\frac{\pi
        \omega}{2 \Sigma}} \overline{u_{\omega,+}^R}}{\sqrt{2\sinh\left(\frac{\pi\omega}{\Sigma}\right)}}
      ,\\
    W_{\omega,-}^{(1)}&=
        \frac{e^{\frac{\pi \omega}{2 \Delta}} u_{\omega,-}^R + e^{-\frac{\pi
        \omega}{2 \Delta}} \overline{u_{\omega,-}^L}}{\sqrt{2\sinh\left(\frac{\pi\omega}{\Delta}\right)}}
        ,
\label{E:W2m}
    &W_{\omega,-}^{(2)}=
        \frac{ e^{\frac{\pi \omega}{2 \Delta}} u_{\omega,-}^L + e^{-\frac{\pi
        \omega}{2 \Delta}} \overline{u_{\omega,-}^R}}{\sqrt{2\sinh\left(\frac{\pi\omega}{\Delta}\right)}}
       ,
\end{align}
are analytic in the lower half imaginary $t$-plane on the
complexified boundary of AdS and that they are normalized to have
Klein-Gordon norm $\pm 1$. Thus, (\ref{E:W1p}) and (\ref{E:W2m}) are
all positive frequency with respect to $t$ on $\partial$AdS${}_3$.
Note that these equations remain valid in the limit $\Delta \to 0$
and $\Sigma \to 0$.

We now introduce creation operators $a_{\omega,\pm}^{(1)\dagger}$,
$a_{\omega,\pm}^{(2)\dagger}$ for the $W$-modes, along with the
corresponding annihilation operators $a_{\omega,\pm}^{(1)}$,
$a_{\omega,\pm}^{(2)}$.  We also introduce creation operators
$b_{\omega,\pm}^{(1)\dagger}$, $b_{\omega,\pm}^{(2)\dagger}$ and
annihilation operators $b_{\omega,\pm}^{(1)}$,
$b_{\omega,\pm}^{(2)}$ associated with the $u$-modes, which are
positive frequency with respect to $\eta$.  The relations between
these operators may be read directly from (\ref{E:W1p}) and
(\ref{E:W2m}):
\begin{align}
a_{\omega,+}^{(1)} &=
        \frac{e^{\frac{\pi \omega}{2 \Sigma}} b_{\omega,+}^R - e^{-\frac{\pi
        \omega}{2 \Sigma}} {b^{L\,\dagger}_{\omega,+}}}{\sqrt{2\sinh\left(\frac{\pi\omega}{\Sigma}\right)}}
        ,
&a_{\omega,+}^{(2)} =
        \frac{e^{\frac{\pi \omega}{2 \Sigma}} b_{\omega,+}^L - e^{-\frac{\pi
        \omega}{2 \Sigma}}
        {b^{R\,\dagger}_{\omega,+}}}{\sqrt{2\sinh\left(\frac{\pi\omega}{\Sigma}\right)}},
        \\
a_{\omega,-}^{(1)} &=
        \frac{ e^{\frac{\pi \omega}{2 \Delta}} b_{\omega,-}^R - e^{-\frac{\pi
        \omega}{2 \Delta}} {b^{L\,\dagger}_{\omega,-}}}{\sqrt{2\sinh\left(\frac{\pi\omega}{\Delta}\right)}}
  ,
&a_{\omega,-}^{(2)} =
        \frac{e^{\frac{\pi \omega}{2 \Delta}} u_{\omega,-}^L - e^{-\frac{\pi
        \omega}{2 \Delta}} {b^{R\,\dagger}_{\omega,-}}}{\sqrt{2\sinh\left(\frac{\pi\omega}{\Delta}\right)}}
        .
\end{align}

Recall that \CFTvacuum is the vacuum on  $\partial$AdS${}_3$. This
means that \CFTvacuum is the minimum energy state with respect to
the time translation $\frac{\partial}{\partial t}$ on
$\partial$AdS${}_3$. As such, it is annihilated by
$a_{\omega,\pm}^{(1)}$, $a_{\omega,\pm}^{(2)}$. We will also be
interested in the state \CFTvacuumeta on $\partial$AdS${}_3$ which
is annihilated by $b_{\omega,\pm}^{(1)}$, $b_{\omega,\pm}^{(2)}$.
The state \CFTvacuumeta induces a state \BTZvacuumeta on
$\partial$BTZ via the identification (\ref{E:identification}).
Because \BTZvacuumeta is annihilated by $b^{(1)}_{\omega, \pm}$,
$b^{(2)}_{\omega, \pm}$ we may identify it as the vacuum state on
$\partial$BTZ.

If we express \CFTvacuum as a set of excitations over
\CFTvacuumeta on $\partial$AdS${}_3$, then  the expression for
\BTZvacuum as a set of excitations over \BTZvacuumeta will follow
immediately. Of course, the expression for  \CFTvacuum as a set of
excitations over \CFTvacuumeta is just the usual Bogoliubov
transformation (see e.g. \cite{BD,Wqft,TJ,SR}):
\begin{equation}
\label{E:vacuum1}
    \CFTvacuum = e^{-i(K_+ + K_-)} \CFTvacuumeta
\end{equation}
with
\begin{equation}
\label{E:Boostedframe}
    K_{\epsilon}=
        i \int_0^{\infty}
        r_{\omega,\epsilon}
        ({b^{R}_{\omega,\epsilon}}^{\dagger}{b^{L}_{\omega,\epsilon}}^{\dagger}-
         {b^{R}_{\omega,\epsilon}} {b^{L}_{\omega,\epsilon}})
         d\omega, \quad \text{and~} \epsilon = + \ {\rm or } \ -,
\end{equation}
where $\tanh(r_{\omega,+}) = e^{-\frac{\pi \omega}{\Sigma}}$ and
$\tanh(r_{\omega,-}) = e^{-\frac{\pi \omega}{\Delta}}$.

It is natural to write
\begin{equation}
\CFTvacuumeta =  \left( |0\rangle_{R+} \otimes |0\rangle_{L+}
\right) \otimes \left( |0\rangle_{R-} \otimes |0\rangle_{L-}
\right) ,
\end{equation}
making use of the decomposition into right- and left-moving modes
($+$ or $-$) supported separately on $D_R$ or $D_L$. We may then
rewrite (\ref{E:vacuum1}) as
\begin{equation}
\label{E:AdSbvacuum}
    |0 \rangle_{b} = \left( e^{-i K_+}|0\rangle_{R,+}|0\rangle_{L,+}\right)
                       \left( e^{-i K_-}|0\rangle_{R,-}|0\rangle_{L,-}\right).
\end{equation}
We see that the right and left movers are entangled with their
partners on the opposite boundary component, but that right-moving
and left-moving particles on the same boundary are not entangled
with each other. Thus, tracing over, say, all left-moving modes
would yield a pure state.

After the identification (\ref{E:identification}), one finds
\begin{equation}
\label{E:BTZbvacuum}
    \BTZvacuum = e^{-i(\tilde{K}_+ + \tilde{K}_-)} \BTZvacuumeta,
\end{equation}
where $\tilde{K}_\pm$ is defined as in (\ref{E:Boostedframe}) but
with the integral over $\omega$ replaced by a sum over the
discrete frequencies $\omega_n = 2\pi n$. We may further write
$\BTZvacuumeta=|0\rangle_L|0\rangle_R$ such that each of the
states $|0\rangle_L$ and $|0\rangle_R$ is the vacuum of a scalar
field on $\mathbb{R}\times S^1$ (i.e., each is a copy of
\CFTvacuum on $\partial$AdS${}_3$). The states $|0\rangle_R$ and
$|0\rangle_L$ are associated respectively with
 $\hat D_L$ and $\hat D_R$.

An examination of (\ref{E:BTZbvacuum}) shows that both the right-
and left- movers are in thermal states, though with different
effective temperatures.  The right movers ($\epsilon=+$) have an
effective inverse temperature $\beta_+ = \frac{2\pi}{\Sigma}$, while
the left movers have an effective inverse temperature $\beta_- =
\frac{2\pi}{\Delta}$. This may also be expressed in terms of the
physical inverse temperature $\beta$ and a chemical potential
$\Omega$ for angular momentum. Such parameters ($\beta, \Omega$) are
related to $\beta_\pm$ through $\beta_\pm = \beta \pm \Omega \beta$.
Thus, we have $\Omega = -\frac{r_-}{r_+}$ and $\beta = \frac{2\pi
\ell r_+}{r_+^2-r_-^2}$.  The quantum state of the zero-modes may be
treated similarly \cite{LouMar}, and again takes the form of a
thermo-field double \cite{TFD} with temperature $\beta = \frac{2\pi
\ell r_+}{r_+^2-r_-^2}$.

Finally, consider the ``high temperature limit'' where either $T_+ =
1/\beta_+ \gg 1$ or $T_- = 1/\beta_- \gg 1$.  As in
\cite{MalStr,LouMar}, one readily shows that the quantum state
\BTZvacuum reproduces the mass, angular momentum, and entropy of the
BTZ black hole in this limit so long as one takes into account the
central charge $c= 6 Q_1Q_5 = 3 \ell/2G_{(3)}$ of the CFT${}_2$ of
\cite{MaldacenaAdSCFT}.  If one also takes into account the
well-known ``fractionization'' effect of the full CFT${}_2$, then
this analysis is valid whenever $T_+ \gg 1/c$ or $T_- \gg 1/c$;
i.e., for all black holes larger than the Planck scale ($r_+ \gg
G_{(3)}$).

\subsection{The extremal limit}
\label{S:extremal}%

Let us now consider the ($M \neq 0$) extremal limit, $r_+ \to
r_-$. Note that the temperature approaches zero while the chemical
potential approaches $-1$ , so that the overall effective
temperature of the right movers remains finite ($T_+ =
\frac{1}{\beta_+} = \frac{1}{(1 + \Omega) \beta}$), while that of
the left movers vanishes ($T_- = \frac{1}{\beta_-} = \frac{1}{(1 -
\Omega) \beta}$).

Our dual description of the extreme black hole remains a state in
the product theory CFT${}_L\otimes$CFT${}_R$ with CFT${}_{R,L}$
associated to $\hat{D}_{R,L}$. Note that CFT${}_R$ is a copy of
the CFT associated with the boundary of AdS${}_3$. On the other
hand, CFT${}_L$ arises from the degenerate $D_L$, which is a
single null line.  Thus, CFT${}_L$ lives on a one-dimensional null
circle and has no left-moving degrees of freedom.  As noted above,
this null circle lies in some sense at the bottom of the infinite
throat of the extreme black hole.  The right-movers of CFT${}_R$
and CFT${}_L$ are entangled in the familiar ``thermo-field double"
state \cite{TFD} at temperature $\beta_+$, while the left-movers
are in their vacuum states.

It is also interesting to consider the BTZ black hole with $M=0$,
obtained by taking $r_+,r_- \rightarrow 0$. We find that the
effective temperature on both boundaries vanishes and that the dual
state is no longer entangled. Instead, we have
$\CFTvacuum=\CFTvacuumeta$ where again $\CFTvacuumeta=| 0 \rangle_L
|0\rangle_R$.  Now, however, $|0 \rangle_L$ is the vacuum of the CFT
corresponding to the point to which $D_L$ collapsed.  Note that
despite the fact that $D_L$ has degenerated to a point, the state
$|0\rangle$ could, in principle, have contained non-tivial
information about the zero-mode on $D_L$.

\section{Discussion}
\label{S:discussion}

In order to investigate the AdS/CFT description of spacetimes with
an infinite throat, we analyzed the dual description of the extreme
BTZ black hole in a simple toy model. At least in the context of our
model, we find that the end of the infinite throat plays a role
analogous to that of a second asymptotic region
\cite{HorMar,MalStr,Maldacena:2001,KOS,entinST,inflate}: the CFT
state dual to an extremal BTZ black hole lives in a product theory
of the form CFT$_R\otimes$CFT$_L$.

Hence, it appears that the eternal extreme black holes may typify a
new class of spacetimes of interest for AdS/CFT. In addition to the
traditional choice of a single asymptotic region resembling the
conformal boundary of AdS${}_3$ (and described by a single CFT), and
also in addition to the case with two such asymptotic regions
studied in
\cite{HorMar,MalStr,LouMar,Maldacena:2001,entinST,inflate,HawMalStr}
(plausibly described by a product of two CFTs), one may also
consider cases with two inequivalent boundary components.  Here we
take one component to be a copy of the boundary of AdS${}_3$, while
the other is a single null circle which must sit at the bottom of
some infinite throat. The suggestion here is that this third class
of boundary conditions may again be associated with a product
CFT${}_L\otimes$CFT${}_R$, where CFT${}_L$ contains only, say,
right-moving degrees of freedom. In such a setting the extreme black
hole may be described as the particular entangled state discussed in
section \ref{S:extremal}.

Further investigation of this idea is certainly needed.  For
example, since the null circle is attached to the bulk in a manner
entirely different from that of the conformal boundary in the
asymptotic region, it is important to study the possible boundary
conditions on this null circle and their influence on the bulk
spacetime.

In addition, it is evident that the relation between boundary
degrees of freedom and those of the bulk will not be as direct as
in the case of conformal boundaries.  In this more familiar
context, at least in the limit where the bulk fields may be
treated semi-classically, one finds
\cite{GKP,Witten9802,BLK,BLKT,R1,R2,R3,Lorentz} that local
operators in the dual CFT are essentially (rescaled) boundary
limits of local bulk operators.  But this seems unlikely to be the
case for our internal infinity, as one may see by studying quantum
field theory on the extreme black hole background.

Consider, for example, a calculation of the bulk state of a linear
quantum field theory on the BTZ background via the same quotient
methods applied to the boundary in section \ref{S:vacuum}.  Note
that this calculation essentially reduces to  calculating the
two-point function $G$, and that $G$ is related to the two-point
function $G_0$ of the vacuum over AdS${}_3$ through a sum over
images.  Furthermore, because this image sum can be performed on the
complexified geometry, analyticity of $G_0$ guarantees that $G$ will
satisfy a KMS condition (see e.g. \cite{KMS1,KMS2}) with respect to
the Killing field which generates the BTZ horizon\footnote{Note that
for the BTZ geometry this Killing field is everywhere timelike in
the bulk (outside the horizon), even in the extreme case $r_+ =
r_-$. This behavior is typical of AdS black holes, and avoids the
issues discussed in \cite{KW} which prohibit the existence of a
Hartle-Hawking state for Kerr black holes in asymptotically flat
spacetimes.}.  As a result, this quantum state will be precisely
thermal, and in particular, mixed with respect to observables
localized in one exterior region.  The thermal ensemble will again
be characterized by the right- and left-moving inverse temperatures
$\beta_\pm$.

Taking the extreme limit, the associated state on the extreme BTZ
background will contain thermal excitations of modes with positive
angular momentum and, as a result, will again be a mixed state.
Thus, studying limits of bulk operators near the end of the
infinite throat will reveal no correlations of the sort entangling
CFT${}_L$ and CFT${}_R$ in our dual CFT state.

One might restate this obseveration more physically by noting that
the Hartle-Hawking state of the non-extreme black hole (with two
asymptotic regions) is an entangled state with respect to modes
localized in each of its asymptotic regions.  These regions are
connected by an Einstein-Rosen bridge.  This bridge becomes
infinitely long in the extreme limit: one side of the bridge
disappears from the spacetime.  Thus, there are no longer any modes
with which to purify the mixed state seen by an observer at the
remaining end of the bridge\footnote{One may always use a
thermofield double construction \cite{TFD} to describe this state as
a pure state living in an extra fictitious Hilbert space. However,
as opposed to the non-extreme case, this extra Hilbert space remains
fictitious and does not have a geometric region in spacetime in
which it may reside.}.  Instead, the complete perturbative bulk
state is mixed for extreme black holes and, from this point of view,
boundary limits of bulk fields do
 not result in the sort of
entanglement described in section \ref{S:extremal}.

A related issue is whether there might be some bulk sense in which
the extreme Hartle-Hawking state can be purified through
entanglement. We leave further investigation of the connection
between our CFT${}_L$ and bulk degrees of freedom for future work.

Finally, although no entanglement is obvious from the bulk
perspective, it is interesting to note that the description of the
CFT as a product CFT${}_L\otimes$CFT${}_R$ is consistent with the
entanglement interpretation of black hole entropy. Such an
interpretation remains mysterious for black holes whose dual lives
in a single CFT but (as emphasized in \cite{entinST}) it becomes
natural if the dual theory takes a product form as above. In
particular, it was advocated in
 \cite{entinST} that the entropy of a
two-asymptotic-region black hole may
always be interpreted as
entanglement entropy of its CFT dual. Here, we see the same
behavior for black holes with one asymptotic region and an
internal infinity.   In particular, we note that \BTZvacuum
encodes the Bekenstein-Hawking entropy through entanglement.

\begin{acknowledgments}
We would like to thank Ofer Aharony, Ramy Brustein, Veronika Hubeny,
Jorma Louko, Mukund Rangamani, and Simon Ross for useful
discussions. Much of this work took place during a visit of A.Y. to
the Kavli Institute of Theoretical Physics at UCSB. A.Y. would like
to thank M. Einhorn and the KITP for providing a stimulating
environment. The work of D.M. is supported in part by NSF grant
PHY0354978 and by funds from the University of California. A.Y. is
supported in part by NSF grant PHY99-07949.
\end{acknowledgments}



\providecommand{\bysame}{\leavevmode\hbox
to3em{\hrulefill}\thinspace}
\providecommand{\MR}{\relax\ifhmode\unskip\space\fi MR }
\providecommand{\MRhref}[2]{%
  \href{http://www.ams.org/mathscinet-getitem?mr=#1}{#2}
} \providecommand{\href}[2]{#2}

\end{document}